\documentstyle[prl,aps,epsf]{revtex}         
\begin{document}
%
\twocolumn[\hsize\textwidth\columnwidth\hsize\csname @twocolumnfalse\endcsname

\title{Suppression and enhancement of impurity scattering in a Bose-Einstein condensate}
\author{ A.P.~Chikkatur, A. G\"{o}rlitz, D.M. Stamper-Kurn,$^{*}$ S. Inouye, 
S. Gupta, and W. Ketterle}
\address{Department of Physics and Research Laboratory of Electronics, \\ 
Massachusetts Institute of Technology, Cambridge, MA 02139}

\date{\today }
\maketitle
\begin{abstract}
Impurity atoms propagating at variable velocities through a trapped 
Bose-Einstein condensate were produced using a stimulated 
Raman transition. The redistribution of momentum
by collisions between the impurity atoms and the stationary
condensate was observed in a time-of-flight analysis.  The
collisional cross section was dramatically reduced when the
velocity of the impurities was reduced below the speed of sound of the
condensate, in agreement with the Landau criterion for
superfluidity.  For large numbers of impurity atoms, we observed 
an enhancement of atomic collisions due to bosonic stimulation. This
enhancement is analogous to optical superradiance.

\end{abstract}
\vskip1pc
]


One manifestation of superfluidity is that objects traveling
below a critical velocity $v_{L}$ through a superfluid propagate without
dissipation. Landau~\cite{land41} used simple kinematic
arguments to derive an expression for the critical velocity
$v_{L}=\min(E(p)/p)$, where $E(p)$ is the energy of an elementary
excitation with momentum $p$.

When superfluid $^{4}$He was forced through capillaries, adsorbed
films and tightly packed powders~\cite{till90}, the onset of
dissipation was found at velocities much lower than
the Landau critical velocity due to turbulence and vortex
formation in the superfluid. The Landau critical velocity can
usually only be observed by moving \emph{microscopic} particles
through the superfluid which do not create a macroscopic
flow pattern.  Studies of superfluidity with
microscopic objects were pursued in liquid $^{4}$He by dragging
negative ions through pressurized
$^{4}$He~\cite{meye61,allu77}, and by scattering $^{3}$He
atoms off superfluid $^{4}$He droplets~\cite{harm99}.

Atomic Bose-Einstein condensates are superfluid gases and show
phenomena analogous to superfluid liquids, albeit at eight orders
of magnitude lower density. For a homogeneous gaseous
Bose-Einstein condensate, the Bogoliubov spectrum indicates a
Landau critical velocity equal to the speed of sound $v_L = c \equiv
\sqrt{\mu/M}$, where $\mu$ is the chemical potential and $M$ is
the mass of condensate atoms.  The first evidence for a critical
velocity in a Bose condensate was obtained by stirring the
condensate with a \emph{macroscopic} object (a laser
beam)~\cite{rama99}.  The observed critical velocity was about a
factor of four smaller than the Bogoliubov speed of sound.
Recent studies of superfluidity have revealed
quantized vortices~\cite{matt99vort,madi00} and a non-classical
moment of inertia~\cite{mara00}.

In this Letter, we report on a study of the motion of \emph{microscopic}
impurities through a gaseous Bose-Einstein condensate. The
impurity atoms were created using a stimulated Raman
process which transferred a small fraction of
the condensate atoms into an untrapped hyperfine state with 
well-defined initial velocity. As these
impurities traversed the condensate, they
dissipated energy by colliding with the stationary condensate, which 
resulted
in a redistribution of momenta of the impurities. As the impurity velocity
was reduced below the speed of sound, we observed a dramatic
reduction in the probability of collisions, which is evidence for
superfluidity in Bose-Einstein condensates.

Our experiments were performed on
Bose-Einstein condensates of sodium atoms in the \mbox{$|F=1,m_{F}=-1\rangle$} 
hyperfine ground state. Condensates of $\sim \!\! 10^{7}$ atoms were 
created using laser and evaporative cooling and stored
in a cylindrically symmetric magnetic trap~\cite{mewe96bec} with an axial
trapping frequency of $16$ Hz. By adiabatically
changing the radial trapping frequency between $165$ Hz and $33$ Hz,
the density of the condensate, and hence the peak speed of sound in the 
condensate was varied between $1.1$ cm/s and $0.55$ cm/s.

Impurity atoms were created using a two-photon Raman transition, in
which the condensate was exposed to a pair of laser beams~\cite{hagl99}.
The laser beams had orthogonal linear polarizations, thus driving a Raman
transition from the trapped $|F=1,m_{F}=-1\rangle$ state to the
untrapped $|F=1,m_{F}=0\rangle$ hyperfine ground state. Both beams
were derived from a common source, and then passed through two
acousto-optic modulators operating with a frequency
difference $\omega=\omega_{z} +
\hbar q^2/(2 M)$, where $\hbar \omega_{z}$ is the Zeeman splitting
between the $|m_{F}=-1\rangle$ and $|m_{F}=0\rangle$ states in the
offset field of the magnetic trap. The momentum transfer from the light field to
the $m_{F}=0$ atoms is $\hbar q = 2 \hbar k
\sin{\theta/2}$ where $k$ is the wavevector of the light field
and $\theta$ is the angle between the two laser beams. The Raman 
light field was typically 
pulsed on for about $10~ \mu s$ at an intensity of several mW/cm$^{2}$. The
fraction of transferred atoms could be varied by changing the
light intensity.

Collisions between the impurities and the condensate were
analyzed by time-of-flight absorption imaging. For this, the
magnetic trap was suddenly switched off 4 ms after the Raman pulse, by
which time the impurity atoms had fully traversed the condensate.
After an additional 5 ms, a magnetic field gradient 
was pulsed on for 30 ms, spatially
separating the $m_{F}=0$ atoms from the condensate. After a
total time-of-flight of typically 60 ms, all atoms were optically
pumped into the \mbox{$|F=2, m_{F}=2\rangle$} ground state and resonantly 
imaged on the cycling transition.

\begin{figure}
        \epsfxsize=60mm
 \centerline{\epsfbox{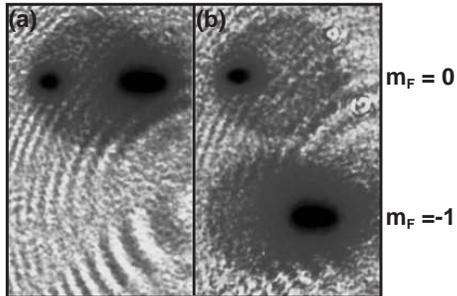}}
    \label{fig1}
\caption{
Observation of $s$-wave halos created by elastic
collisions between the condensate and impurity atoms traveling at
6 cm/s (to the left in
images). (a) Absorption image
after 50 ms of time-of-flight shows the velocity distribution
after collisions between the condensate and the outcoupled
$m_{F}=0$ atoms. The collisional products are distributed
over a sphere in momentum space. (b) same as a), but using a
Stern-Gerlach type magnetic field gradient to separate the
$m_{F}\!=\!\!0$ atoms (top) from the $m_{F}\!=\!\!-1$ condensate 
(bottom). The fringes are an imaging artifact.
Images are 4.5 $\times$ 7.2 mm}
\end{figure}

Collisions at ultracold temperatures are in the $s$-wave regime.
The products of such collisions between free particles are
evenly distributed in momentum space over a spherical shell
around the center-of-mass momentum of the collision partners.
A time-of-flight picture records the momentum distribution
of the released cloud. Thus, collisions between the condensate and the
impurities are visible as a circular halo which
represents the line-of-sight integrated spherical shell.  Fig.~1 
shows a
typical absorption image of collisions in the free particle regime 
for impurity atoms with a velocity of $2 \hbar k/M$ = 6 cm/s, 
produced by counterpropagating Raman beams.

To probe for superfluidity, we produced impurity atoms at low
velocities (7 mm/s) by using Raman beams which intersected at an 
angle of
$\approx \! \! 14^{\circ}$ and aligned symmetrically about
the radial direction, so that the difference vector ${\bf q}={\bf
k_{1}-k_{2}}$ was directed axially~\cite{stam99phon}.  The
trajectory of the impurity atoms was initially in the axial
direction, but was soon modified  by two forces: a
downward gravitational acceleration along a radial direction
(into the page in images and hereafter denoted as the $z$-axis), and the
radial
mean-field repulsion of the $m_F = 0$ atoms from the $m_F = -1$
condensate.  

This situation is similar to the previous study of an rf
output coupler by which $m_F = 0$ atoms were produced at rest
\cite{mewe97}.  For the rf
output coupler, collisions between impurity atoms and the condensate were
difficult to detect because the scattered and unscattered
atoms were not clearly distinguishable.  In this study, the small axial 
velocity imparted by Raman scattering allowed us to identify products of
elastic collisions in time-of-flight images since
collisions with the stationary
condensate tended to redistribute the impurity atoms toward lower
axial velocities. However, 
the acceleration of
the impurity atoms precluded the observation of well-defined collision
halos.
A time-of-flight analysis of impurity
scattering for the case of a low density condensate is shown in
Figure 2a.  The axial velocity imparted by Raman scattering displaces the
unscattered $m_F = 0$ atoms upward in the image, whereas collisions
produce impurity atoms with smaller axial velocities which then appear
below the unscattered atoms in the image.  In contrast, Figure 2b shows a
time-of-flight image for the case of a high density condensate, for which the
number of collided atoms is greatly
diminished, indicating the suppression of impurity collisions due to
superfluidity.

\begin{figure}
    \epsfxsize=70mm
 \centerline{\epsfbox{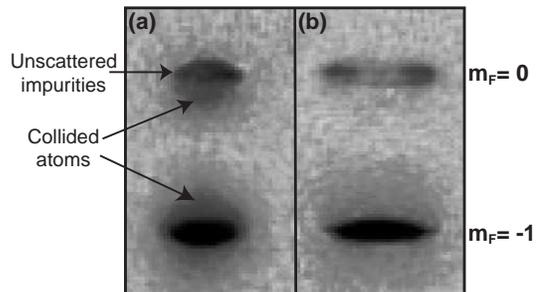}}
    \label{fig2}
\caption{
Superfluid suppression of collisions.  The impurity
$m_{F}=0$ atoms (top) traveled at 7 mm/s along the condensate axis 
(upward in image)
and were separated from the condensate (bottom) by a 
magnetic field gradient applied during ballistic expansion. (a) Absorption
image after 50 ms of time of flight shows the collisional products
as indicated by the arrow. For
this image, $v_{g}/c =2.7$ (see text). (b) Similar image as (a) with 
$v_{g}/c =1.6$. Collisions are suppressed. The outcoupled atoms were
distorted by mean-field repulsion.  The images are 2.0 $\times$
4.0~mm. }
\end{figure}
The number of collided atoms was determined by counting impurity atoms in
a region of the time-of-flight image below the unscattered impurity
atoms, which also contained Raman outcoupled thermal 
$m_F = 0$ atoms. Thus, the number of 
collided atoms in the counting region was obtained by subtracting the thermal
background which was determined by counting a similar sized region above
the unscattered impurity atoms where we expect few collision
products.  This number was doubled to obtain the total number of collided atoms
since we expect only about half of the collision products to be in the counting
region; the remainder overlapped with the distribution of unscattered
impurity atoms.

\begin{figure}
   \epsfxsize=50mm
 \centerline{\epsfbox{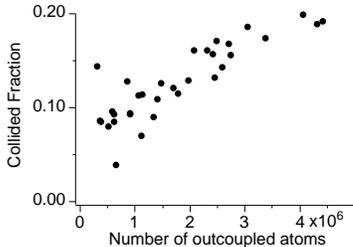}}
\caption{ Collective amplified elastic scattering in a
Bose-Einstein condensate. Shown is the fraction of collided atoms
vs.\ the number of outcoupled atoms. For this data,  $v_{g}/c$=4.9 and
the chemical potential was 1.8 kHz.}
 \label{ampli}
\end{figure}

In studying these collisions, we discovered that the fraction of
collided atoms increased with the number of outcoupled impurities
(see Fig.~\ref{ampli}). According to a perturbative 
treatment described below, the collision probability should be
independent of the number of impurities.  If the number of
outcoupled atoms is increased, one would expect the
collision probability to \emph{decrease} slightly due to the
reduction in the condensate density, or to
\emph{increase} slightly because the smaller condensate
density implies a smaller critical velocity for dissipation.
However, these effects are smaller (10--20\%) than the observed 
two-fold increase in the collided fraction.

Rather, this large increase can
be explained as a collective self-amplification of atomic
scattering, akin to the recently observed superradiant
amplification of light scattering from a Bose-Einstein condensate
\cite{inou99super}. Collisions between impurity atoms and the
condensate transfer atoms from a macroscopically occupied initial
state to final momentum states which were previously empty.  The
population in these final states can stimulate further
scattering by bosonic enhancement and this effect increases for larger
outcoupling. This collisional
amplification is not directional, and is similar to the recently
observed optical omnidirectional superfluoresence~\cite{lvov99}.
In contrast, the observation of four-wave mixing of atoms~\cite{deng99} 
represents the case where collisions were stimulated
by a single macroscopically occupied final mode.

\begin{figure}
    \epsfxsize=50mm
 \centerline{\epsfbox{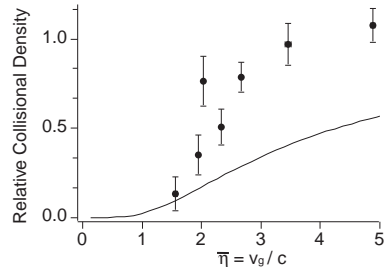}}
\caption{Onset of superfluid suppression of collisions.  Shown is
the observed collisional density normalized to the predicted collisonal
density ${\mathcal C}_{\infty}$ in the limit of high velocities as
the density of the condensate was increased (plotted from right to left).
The collisional density is proportional to the collision probability
of the outcoupled atoms (see text). The $x$-axis shows
$\overline{\eta}=v_{g}/c$ which is a measure of the
impurity velocity over the speed of sound.
The solid line is the theoretically
expected collisional density calculated by a numerical integration
of Eq.~\ref{collint}. The error bars represent the
statistical uncertainty.}
 \label{fig4}
\end{figure}

Fig.~\ref{fig4} shows the decrease of collision probability as the 
velocity of the impurity
atoms approached the speed of sound in the condensate. The 
collision probability was determined by
averaging over many iterations of the experiment with the
number of outcoupled atoms kept below $10^6$, in which case
collective effects may be neglected. 
For our experimental conditions, the impurity velocity 
was predominantly determined by the gravitational
acceleration $g=9.8$ m/s$^{2}$, which imparted an average velocity of
$v_{g}=\sqrt{2 g z_{c}}$ where $z_{c}$ is the Thomas-Fermi radius
of the condensate in the $z$-direction.  
This downward
velocity ranged from 17 mm/s for tightly confined condensates to
26 mm/s for loosely confined condensates, and was larger than the
initial 7 mm/s velocity imparted by Raman scattering.  Thus, the effect
of superfluidity on impurity scattering depends primarily on
the parameter $\overline{\eta} = v_g / c$ which is the ratio of the
typical impurity velocity $v_g$ to the speed of sound at the
center of the condensate $c = \sqrt{\mu / M}$. 
Experimentally,
$\overline{\eta}$ is determined using the radial trapping frequency
and the
chemical potential $\mu$ which is determined from the expansion of the 
condensate in the time-of-flight images \cite{kett99var}.

High and low $\overline{\eta}$ correspond to dissipative and 
non-dissipative regimes, respectively.
This
behavior is apparent in Fig.\ 2.  For the case of loose
confinement, the low condensate density (small $c$) and large
condensate radius (large $v_g$) yield a large value for
$\overline{\eta} \sim 5$.  The effect of collisions is clearly
visible with about 20\% of the atoms scattered to lower axial
velocities (Fig.\ 2a). In contrast, for the case of tight
confinement, the high condensate density (large $c$) and small
condensate radius (small $v_g$) yield a small value for
$\overline{\eta} \sim 1.5$, and the collision probability is greatly
suppressed due to superfluidity (Fig.\ 2b).

The predicted cross-section for collisions between an $m_F=0$ impurity
atom at momentum $\hbar \bf{k}$ and a $m_F = -1$ condensate of density
$n_0$ is obtained by calculating the collision rate $\Gamma$ using Fermi's
Golden rule~\cite{timm98super,idzi99}:
\begin{eqnarray*}
 \Gamma
    & = & n_{0}\left(\frac{\hbar a}{M}\right)^{2}
    \int \!\! dq d\Omega \,\, q^{2} S(q) \,\,
	\delta \left(\frac{\hbar {\bf k \cdot
    q}}{M}-\frac{\hbar q^{2}}{2 M}- \omega_{q}^{B} \right) \nonumber \\
	&=& n_{0} \: \sigma(\eta) \: v, \nonumber
\end{eqnarray*}
Here, $S(q) = \omega_q^0 / \omega_q^B$ is the
static structure factor
of the condensate with $\hbar \omega_q^0 = \hbar^2 q^2 / 2 M$ and $\hbar
\omega_q^B = \sqrt{ \hbar \omega^{0}_{q} (\hbar \omega^{0}_{q} + 2\mu)}$
being the energies of a free particle and a Bogoliubov quasiparticle
of momentum $q$, respectively.  The collision
cross section is $\sigma(\eta) = \sigma_0 F(\eta)$ where $\eta =
v/c$, $v = \hbar k / M$ is the impurity velocity, and 
$\sigma_0 = 4 \pi a_{0,-1}^2$ where
$a_{0,-1} = 2.75$ nm~\cite{ties96} is the scattering length
for $s$-wave collisions between 
the \mbox{$|m_{F}=0 \rangle$} and \mbox{$|m_{F}=-1\rangle$} states of sodium.
For $\eta<1$, $F(\eta)=0$ and for $\eta>1$,
$F(\eta)=1-1/\eta^{4}-\log(\eta^{4})/\eta^{2}$. 

We can approximate our experiment 
by considering the motion of the 
$m_F=0$ atoms under the gravitational acceleration alone and ignoring 
the effects of the initial axial velocity and 
mean-field expulsion~\cite{meanfield_footnote}. The
$m_{F}=0$ atoms falling through the condensate experience
a collisional density ${\mathcal{C}}(\eta)=\int\!\!{dz}
\: n(x,y,z) \sigma(\eta)$, where
$n(x,y,z)$ is the condensate
density, and $\eta$ is determined by the local
condensate density and the downward impurity velocity. The collisional
density relative to its value at large velocities 
${\mathcal{C}}_{\infty}$is given by 
\begin{eqnarray}
    \frac{{{\mathcal C}}(\overline{\eta})}{{\mathcal C}_{\infty}} 
    & \approx &
    \frac{\int\!\! d{\bf r} \; n_{I}({\bf r})\times 
 \int\!\! dz' \; n({\bf r}') \; \sigma_{0} \; F(\eta)}
 {\int\!\! d{\bf r} \; n_{I}({\bf r})\times 
 \int\!\!dz' \; n({\bf r}') \; \sigma_{0}}
\label{collint}
\end{eqnarray}
where we assume that the initial impurity density 
$n_I({\bf r})\!\!\propto \!\! n({\bf r})$~\cite{copy_footnote}. 
The condensate density in the Thomas-Fermi
limit is
$n(x,y,z)=n_{0}(1-(x/x_{c})^{2}-(y/y_{c})^{2}-(z/z_{c})^{2})$,
where $x_{c}=(2 \mu / M \omega_{x}^{2})^{1/2}$ (similarly for
$y_{c}$ and $z_{c}$) is the Thomas-Fermi radius, $\omega_{x}$ is
the trapping frequency in the $x$ direction, and $\mu$ is the
chemical potential.  The solid line in Fig.~\ref{fig4} was
determined by numerically integrating Eq.~\ref{collint}.  To 
compare the collision probability for the different data points, we 
divided the observed collided fraction by 
${\mathcal C}_{\infty}=(5/12) \times n_{0} \sigma_{0} z_{c}$. The
observed decrease in the collisional
density for small $\overline{\eta}$ (Fig.~\ref{fig4}) shows the
superfluid suppression of collisions. Numerical simulations 
ruled out the possibility that the observed decrease in collisional 
density could be caused solely by variations of the path length of particle 
trajectories 
due to the mean-field repulsion and the initial velocity.

The measured values in Fig.~\ref{fig4} are 
systematically larger by a factor of about two than those expected 
theoretically. This discrepancy is also seen for 
impurity collisions at velocities of 6 cm/s for which superfluidity 
should play no role. While we cannot presently account for this
systematic error, the observation of suppression of collisions due to 
superfluidity is robust, since it
requires only a relative comparison of collision probabilities at
different $\overline{\eta}$. 

The method presented here can generally be used to study ultra-cold collisions.
In this study,
we focused on collisions between
atoms in different hyperfine states. By driving a Bragg
transition~\cite{sten99brag} instead of a Raman
transition, we have also observed collisions between atoms in the 
same internal state.
At a velocity of 6 cm/s,
we found the collision cross section to be $2.1\pm 0.3$ times 
larger than in the Raman
case, reflecting the exchange term in elastic collisions for identical
particles that increases the cross section from
$4 \pi a^2$ to $8 \pi a^2$.

Raman transitions are one way to realize output couplers for atom
lasers~\cite{hagl99,mewe97,bloc98,ande98atla}. 
Theoretical treatments of atom lasers have typically
considered only the condensate and the outcoupled atoms in a
two-mode approximation and 
ignored the modes accessible by collisions~\cite{atomlaser}. However,
our experiment shows
that the outcoupled atoms do not simply pass through the
condensate. Rather, they collide and populate modes coupled by 
atomic scattering~\cite{band99}; the collisions may even be enhanced by bosonic
stimulation. In principle, such
collisional losses can be avoided by \emph{lowering} the density.
However, an alternative route to suppressing
collisions is to \emph{increase} the density until the speed of
sound is larger than the velocity of the outcoupled atoms, thus
realizing a ``superfluid'' output coupler. 

In conclusion, we have studied collisions between impurity atoms and a Bose-Einstein condensate.
Both the observed superfluid suppression of collisions and the collective enhancement are crucial
considerations
for the future development of intense atom lasers.

We are grateful to D.E. Pritchard for valuable discussions.
This work was supported by the ONR, NSF, JSEP, ARO, NASA, and the
David and Lucile Packard Foundation. A.P.C. 
acknowledges additional support from the NSF, A.G. from DAAD, and
D.M.S.K. from JSEP and a Robert A. Millikan Postdoctoral
Fellowship.


\begin{thebibliography}{10}
\bibitem{meanfield_footnote}
A numerical simulation showed that including these effects changes the
  collisonal density by less than 20\%.
\bibitem{copy_footnote}
The Raman transition strength is uniform over the entire condensate since the
  frequency broadening due to finite pulse-length (16 kHz) was larger than the
  broadening due to mean-field ($\sim 2$ kHz).
\end{thebibliography}
\end{document}